\documentclass[10pt,aps,prb,twocolumn, nofootinbib]{revtex4-1}
\usepackage{amssymb,amsmath}
\usepackage{subfig}
\usepackage{graphicx}
\usepackage{color}
\usepackage{hyperref}
\usepackage{listings}
\usepackage[utf8]{inputenc}
\usepackage{booktabs}
\usepackage{multirow}
\usepackage{array}
\usepackage{tabu}
\usepackage{float}
\usepackage{varioref}
\hypersetup{
    unicode=false,          
    pdftoolbar=true,        
    pdfmenubar=true,        
    pdffitwindow=false,     
    pdfstartview={FitH},    
    pdfauthor={Kevin Multani},     
    colorlinks=true,       
    linkcolor=blue,          
    citecolor=blue,        
    urlcolor=blue         
}
\graphicspath{{figures/}}

\begin{document}

\definecolor{dkgreen}{rgb}{0,0.6,0}
\definecolor{gray}{rgb}{0.5,0.5,0.5}
\definecolor{mauve}{rgb}{0.58,0,0.82}

\lstset{frame=tb,
  	language=Matlab,
  	aboveskip=3mm,
  	belowskip=3mm,
  	showstringspaces=false,
  	columns=flexible,
  	basicstyle={\small\ttfamily},
  	numbers=none,
  	numberstyle=\tiny\color{gray},
 	keywordstyle=\color{blue},
	commentstyle=\color{dkgreen},
  	stringstyle=\color{mauve},
  	breaklines=true,
  	breakatwhitespace=true
  	tabsize=3
}

\title{\textbf{Engineering of spin mixing conductance in Ru/FeCo/Ru interfaces: Effect of Re Doping}}
\author{Rahul Gupta\hyperref[itm:Rahul]{$^1$}, Nilamani Behera\hyperref[itm:Rahul]{$^1$}, Vijay A. Venugopal\hyperref[itm:VJ]{$^2$}, Swaraj Basu\hyperref[itm:VJ]{$^2$}, Mark A. Gubbins\hyperref[itm:VJ]{$^2$}, Lars Bergqvist\hyperref[itm:Lars]{$^{3,4}$}, Rimantas Brucas\hyperref[itm:Rahul]{$^1$}, Peter Svedlindh\hyperref[itm:Rahul]{$^1$}\hyperref[itm:corrs]{$^*$}, Ankit Kumar\hyperref[itm:Rahul]{$^1$}\hyperref[itm:corrs]{$^*$}
}

\affiliation{\label{itm:Rahul}{$^1$}{Department of Engineering Sciences, Uppsala University, Box 534, SE-751 21 Uppsala, Sweden}\\ 
\label{itm:VJ}{$^2$}Seagate Technology, BT47 6SE, Londonderry, United Kingdom,\\
\label{itm:Lars}{$^3$}Department of Applied Physics, School of Engineering Sciences, KTH Royal Institute of Technology, Electrum 229, SE-16440 Kista,\\ 
\label{itm:Lars}{$^4$}Sweden Swedish e-Science Research Center, KTH Royal Institute of Technology, SE-10044 Stockholm, Sweden\\ 
}

\date{\today}

\begin{abstract}
\section*{Abstract}
We have deposited polycrystalline Re doped $(Fe_{65}Co_{35})_{100-x}Re_{x}$ (0 $\leq$ x $\leq$ 12.6 at\%) thin films grown under identical conditions and sandwiched between thin layers of Ru in order to study the phenomenon of spin pumping as a function of Re concentration. In-plane and out-of-plane ferromagnetic resonance spectroscopy results show an enhancement of the Gilbert damping with an increase in Re doping. We found evidence of an increase in the real part of effective spin mixing conductance [Re($g^{\uparrow\downarrow}_{eff}$)] with the increase in Re doping of 6.6 at\%, while a decrease is evident at higher Re doping. The increase in Re($g^{\uparrow\downarrow}_{eff}$) can be linked to the Re doping induced change of the interface electronic structure in the non-magnetic Ru layer and the effect interfacial spin-orbit coupling has on the effective spin-mixing conductance. The lowest and highest values of Re($g^{\uparrow\downarrow}_{eff}$)  are found to be 9.883(02) $nm^{-2}$ and 19.697(02) $nm^{-2}$ for 0 at\% and 6.6 at\% Re doping, respectively. The saturation magnetization decreases with increasing Re doping, from 2.362(13) T for the undoped film to 1.740(03) T for 12.6 at\% Re doping. This study opens a new direction of tuning the spin-mixing conductance in magnetic heterostructures by doping of the ferromagnetic layer, which is essential for the realization of energy efficient operation of spintronic devices.

\label{itm:corrs}{$^*$}corresponding author: ankit.kumar@angstrom.uu.se, peter.svedlindh@angstrom.uu.se

\end{abstract}

\maketitle
\section{\label{sec:one}Introduction}
Generation, transportation and detection of pure spin currents play a fundamental role in spintronic devices. Pure spin currents can be generated in ferromagnet and normal metal (FM/NM) bilayer structures, which are governed by the magnetization dynamics via spin pumping [\onlinecite{PhysRevLett.88.117601}]. With the uniform ferromagnetic resonance (FMR) mode excited in the FM/NM structure, transfer of spin angular momentum occurs from the FM layer to the NM layer via the interface, due to magnetization precession. This phenomenon is called spin pumping, which can be measured by the FMR technique [\onlinecite{PhysRevB.19.4382,PhysRevB.66.104413,PhysRevLett.97.216603,saitoh2006conversion}]. Quantitatively, the spin pumping is determined by the spin mixing conductance (SMC) and damping ($\alpha$) of the spin dynamics [\onlinecite{PhysRevB.66.224403},\onlinecite{brataas2004spin}].

The generalized form of the pure spin current (a 2 $\times$ 2 matrix in spin space) is governed by the spin-dependent conductance, which depends on the reflection (r) and transmission (t) matrices. The SMC is defined as $g^{\uparrow\downarrow}$ =  $\frac{e^2}{h} [M-\Sigma r^\uparrow (r^\downarrow)^*]$, where $M$ is the number of propagating quantum channels at the Fermi level and $r^{\uparrow(\downarrow)}$ is the reflection matrix for spin-up (spin-down) electrons. The first part of this equation $(\frac{e^2}{h} M)$ is called the Sharvin conductance, where M is proportional to the Fermi surface-averaged density of states at the interface [\onlinecite{li2018effect}]. The SMC is a complex parameter $g^{\uparrow\downarrow} = Re(g^{\uparrow\downarrow}) + \iota 
Im(g^{\uparrow\downarrow})$ [in units of $\Omega^{-1}$ $m^{-2}$ (or per unit quantum conductance per unit area)], where $Re(g^{\uparrow\downarrow})$ and $Im(g^{\uparrow\downarrow})$ are the real and imaginary parts of the SMC, respectively [\onlinecite{PhysRevB.66.224403}]. However, for realistic interfaces, as per first principle calculations, the imaginary part of $g^{\uparrow\downarrow}$ is negligible for metallic systems in the absence of interfacial Rashba states [\onlinecite{PhysRevB.65.220401}]. This approximation has been considered throughout the paper. The SMC also varies between different combinations of FM/NM layers [\onlinecite{gerrits2006enhanced,carva2007ab,PhysRevLett.107.046601,deorani2013role,weiler2014detection}]. For example, $g^{\uparrow\downarrow}$ is found to be  about six times smaller for low conductivity FMs (\textit{e.g}. $Fe_3O_4$) than for highly conductive FMs (3d transition metal alloys; \textit{e.g.} permalloy and Heusler compounds) with Pt as NM layer [\onlinecite{PhysRevLett.107.046601}]. Moreover, M. Schoen  \textit{et. al.} [\onlinecite{schoen2016ultra}] (in supplementary information) reported the compositional dependence of the SMC in iron-cobalt (FeCo) alloy thin films. Studies of FeCo have led to many advances in the fundamental understanding of magnetic devices because of its metallic nature and ultra-low damping [\onlinecite{schoen2016ultra}], providing a low threshold current density in spin-transfer-torque magnetic-random-access-memories. However, interfacial properties of FeCo based heterostructures are still to be investigated in detail. In addition, engineering the properties of FM layer and its interface with NM layer are required to achieve high efficient operation of spin-logic devices, e.g., spin torque MRAM [\onlinecite{PhysRevApplied.5.014002}]. The next generation spin torque MRAM for high-density catch memory requires high magnetic anisotropy for thermal stability and high damping for ultrafast operation [\onlinecite{6374706},\onlinecite{bhatti2017spintronics}]. The benchmark for efficient operation of spin-logic devices is the product of the switching power and the delay, where the delay represents the switching time. The power-delay product describes the average energy consumed per switching event, and hence balances the performance and energy consumption. However, there is trade-off between ultrafast switching and energy efficient operations as an increase in damping enhances the writing energy. A reduction of the power-delay can be achieved by enhancing the SMC of the FM/NM interface.
The energy-delay reduction scaling can be achieved by enhancing the SMC of the FM/NM layers interface. The SMC governs the efficiency of spin angular momentum transfer across the interface, providing spin torques to switch the magnetic state of the FM layer in MRAMs. Manipatruni \textit{et al.} have reported that by increasing the SMC at the FM/NM interface, the power-delay product of in-plane all spin-logic devices can be reduced significantly [\onlinecite{PhysRevApplied.5.014002}].

Doping the FM layer with 5d transition elements can be used to tune the magnetic properties [\onlinecite{PhysRevB.87.014430}], providing a way of optimizing  magnetic heterostructures for efficient spin logic devices. For example, it is expected that damping relaxation due to two-magnon-scattering (TMS) increases with non-magnetic doping. Moreover, other contributions to the damping, such as radiative and eddy current contributions, owing to the inductive coupling between the sample and the co-planar waveguide (CPW), also have to be considered when estimating the SMC of the heterostructures; the extrinsic contributions have to be subtracted from the total measured damping in order to reliably extract the interfacial properties.

To tune the interfacial and dynamic properties, as per theory and experimental suggestions [\onlinecite{schoen2016ultra}], we used low damping $Fe_{65}Co_{35}$ thin films with Ru as seed and capping layers, as Ru can help to decrease the coercivity and the effective damping parameter of $Fe_{65}Co_{35}$ thin films [\onlinecite{akansel2018effect}]. In the present work, by 6.6 at\% Re-doping of technologically important $Fe_{65}Co_{35}$ interfaced with Ru layers, we have evidenced a 100\% increase of the SMC along with 135\% increase of the damping, while the saturation magnetization decreases marginally. These properties make Re-doped $Fe_{65}Co_{35}$ thin films interesting for ultrafast and energy efficient spin torque logic devices.

\begin{figure}[b]
\includegraphics[width=10cm]{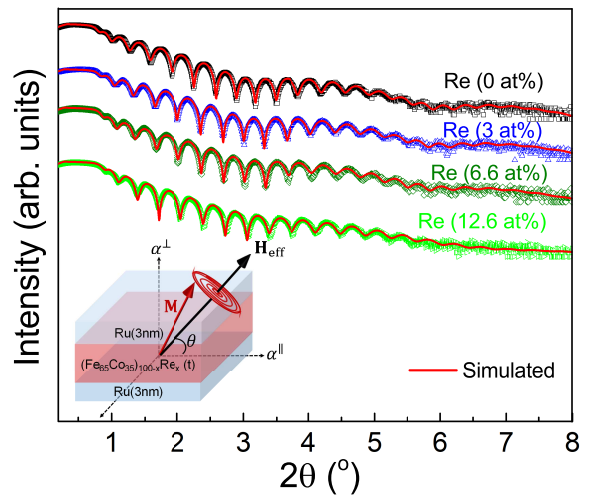}\\
\caption{XRR spectra of FeCo(20 nm) for all Re doping concentrations. Red lines show the simulated spectra and open symbols represent observed data. Inset is a schematic of our sample configuration.}\label{fig:fig1}
\end{figure}

\section{\label{sec:two}EXPERIMENTAL DETAILS}
Using DC magnetron sputtering, rhenium (Re) doped polycrystalline $Fe_{65}Co_{35}$ alloy thin films with different thickness were deposited at room temperature on $Si(100)/SiO_2$ substrates with Ru as seed and capping layers. The Re concentration (x) in the $(Fe_{65}Co_{35})_{100-x}Re_{x}$ (0 $\leq$ x $\leq$ 12.6 at\%) thin films was varied by changing the deposition rate of Re. The target power, base pressure and Ar working pressure in the vacuum chamber were the same during all the depositions; for further details see Ref. [\onlinecite{akansel2019enhanced}]. The nominal thicknesses of the FeCo films was 5, 10, 15, 20 and 30 nm and the nominal thickness of the Ru layer was 3 nm. The sample configuration is shown in the inset of Fig. \hyperref[fig:fig1]{1}.

Our polycrystalline films have the body-centered cubic crystal structure. The lattice constant increases with increasing Re doping, which follows the trend from theoretical calculations [\onlinecite{akansel2019enhanced}]. Rutherford backscattering spectrometry was used to discover the absolute elemental composition of Co, Fe, Ru and Re in our films. The detailed analysis can be found in our previous work [\onlinecite{akansel2019enhanced}].

Film thickness, surface/interface roughness, and density were measured by X-ray reflectivity (XRR) using the X’Pert-Pro system. The scan angle (2$\theta$) was set to 0-8$^\circ$ during the measurements. Fittings were done in specular mode using the HighScore software. The XRR spectra of the 20 nm thick of FeCo films for all Re concentrations are shown in Fig. \hyperref[fig:fig1]{1}. Resistivity measurements were performed in a Quantum Design Physical Property Measurement System (PPMS) using the four-probe method.

The spin dynamic properties were studied using IP- and OP-FMR spectroscopy. The direction of the external magnetic field was parallel and perpendicular to the film plane in IP- and OP-FMR geometry, respectively, as shown in the inset of Fig. \hyperref[fig:fig1]{1}. The IP- and OP-FMR measurements were based on lock-in amplifier and vector-network-analyzer techniques, respectively. The samples (4 $\times$ 4 mm$^2$ in size) were mounted face-down on the CPW and a fixed frequency ($f$) and amplitude microwave field was passed through the CPW using a microwave source [\onlinecite{PhysRevB.96.224425}]. The measurements were performed in field-sweep mode; the uniform FMR resonance mode was observed in the presence of the magnetic field, fulfilling the resonant condition for the material. The resonance field will hereafter be referred to as $H_r$ and the linewidth of the resonance will be denoted $\Delta H$.

\begin{table*}
\caption{A comparison of nominal ($t_{FeCo}^N$) and exact ($t_{FeCo}^E$) thickness (in nm) of FeCo thin films from XRR analysis. $g^{\perp}$ and $\mu_0M_{eff}$ (in T) are fitted values from OP-FMR analysis. Errors are mentioned in within paranthesis, which correspond to the last digit of the values.}
\label{tab:table1}
\begin{tabular}{c | c c c | c c c | c c c | c c c}
\hline
\hline
\textbf{} & \multicolumn{3}{ c |}{\textbf{Re (0 at\%)}} & \multicolumn{3}{ c |}{\textbf{Re (3 at\%)}} & \multicolumn{3}{ c |}{\textbf{Re (6 at\%)}} & \multicolumn{3}{ c }{\textbf{Re (12.6 at\%)}} \\
$t_{FeCo}^N$ & $t_{FeCo}^E$ & $g^{\perp}$ & $\mu_0M_{eff}$ & $t_{FeCo}^E$ & $g^{\perp}$ & $\mu_0M_{eff}$ & $t_{FeCo}^E$ & $g^{\perp}$ & $\mu_0M_{eff}$ & $t_{FeCo}^E$ & $g^{\perp}$ & $\mu_0M_{eff}$ \\
\hline
5 & 5.08(02) & 2.044(03) & 2.055(01) & 4.53(03) & 2.044(04) & 1.877(01) & 4.47(03) & 2.020(02) & 1.667(01) & 3.57(03) & 2.049(10) & 1.274(04)\\
10 & 9.58(02) & 2.036(05) & 2.202(02) & 9.03(05) & 2.098(04) & 2.038(01) & 8.78(05) & 2.041(03) & 1.839(01) & 9.80(05) & 2.065(01) & 1.484(06)\\
15 & 13.16(02) & 2.066(05) & 2.265(01) & 13.52(07) & 2.045(04) & 2.079(01) & 13.29(07) & 2.056(08) & 1.902(03) & 13.18(08) & 2.022(11) & 1.544(04) \\
20 & 19.41(01) & 2.065(01) & 2.288(007) & 18.25(01) & 2.057(06) & 2.101(02) & 18.64(02) & 2.023(08) & 1.883(03) & 16.09(03) & 2.076(009) & 1.609(003)\\
30 & 28.35(01) & 2.014(03) & 2.277(01) & 26.55(02) & 2.075(01) & 2.155(003)  & 26.76(02) & 2.039(03) & 1.940(01)& 26.25(01) & 2.031(03) & 1.620(01)\\
\hline
\hline
\end{tabular}
\end{table*}

\begin{figure}[b]
\centering
\includegraphics[width=8cm]{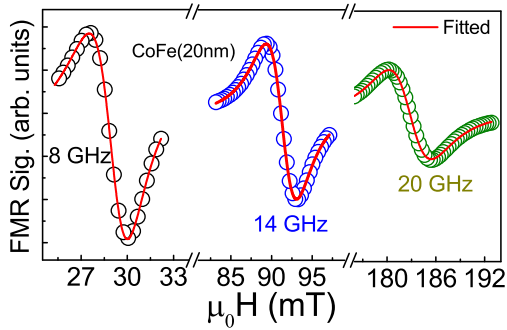}\\
\caption{Typical absorption spectra from IP-FMR of FeCo(20 nm) without Re doping at 8, 14 and 20 GHz frequencies. Open symbols and red lines are observed and fitted data, respectively.}\label{fig:fig2}
\end{figure}

\begin{figure}[b]
\includegraphics[width=9.5cm]{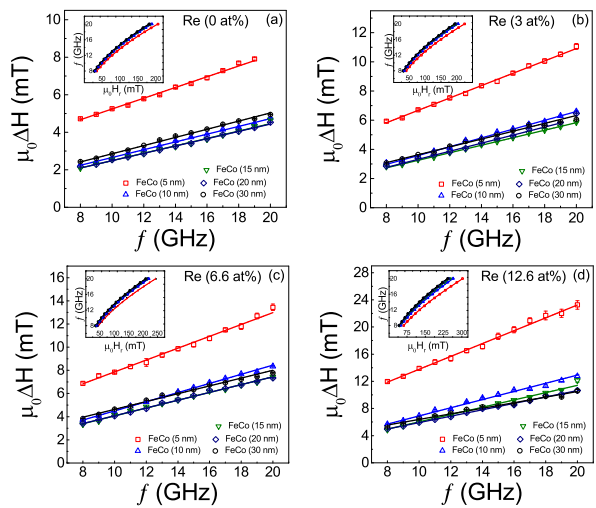}\\
\caption{$\mu_0$ $\Delta H$ vs. $f$ from IP-FMR. The insets correspond to $f$ vs. $\mu_0$ $H_r$. Open symbols and lines are experimental and fitted data, respectively.}\label{fig:fig3}
\end{figure}

\section{\label{sec:three} Result and Discussion}
Results from XRR measurements provide accurate estimation of thickness, roughness and density of the different layers in the $SiO_2/Ru/FeCo(t_{FeCo})/Ru/Oxide$ samples. Fig. \hyperref[fig:fig1]{1} shows observed (open symbols) along with simulated spectra (red lines) for FeCo (20 nm) with different Re concentrations; the results from the analysis are presented in Table \hyperref[tab:table1]{I}. The model used for precise fitting, $Ru/(Fe_{65}Co_{35})_{100-x}Re_x/Ru/Oxide$, includes 4 layers, where the FeCo layer can have different Re concentration. We use the XRR determined thicknesses of the FeCo layers for further calculations, represented by $t_{FeCo}^E$.

\subsection{IP-FMR}

\begin{figure}[b]
\centering
\includegraphics[width=7cm]{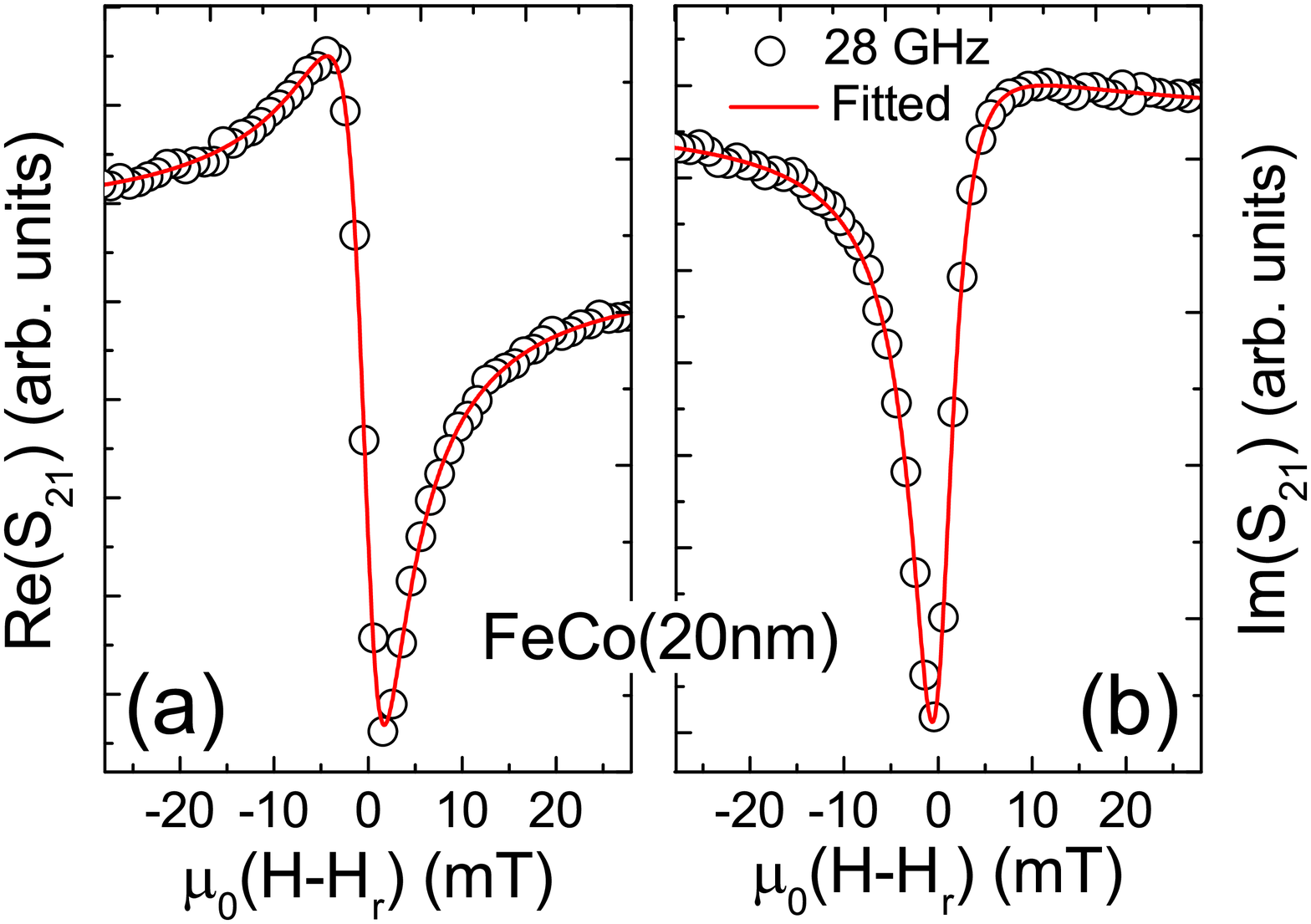}\\
\caption{Re$(S_{21})$ and Im$(S_{21})$ vs. $\mu_0(H-H_r)$ of undoped FeCo(20 nm) at 28 GHz. Black open symbols and red lines are experimental and fitted data, respectively.}\label{fig:fig4}
\end{figure}

To understand the fundamental behavior of the magnetization dynamics, IP-FMR spectra were recorded using a signal generator–detector diode based lock-in technique. The microwave frequency range used in these measurements was from 8 to 20 GHz. Fig. \hyperref[fig:fig2]{2} shows typical IP-FMR spectra of a FeCo(20 nm) sample at different frequencies. To find the resonance field and linewidth (half width at half maximum, HWHM), the observed IP-FMR spectra were fitted with the field derivative of the susceptibility function [$\chi(H,t)$] as given below [\onlinecite{woltersdorf2004spin}],

\footnotesize 
\begin{equation}
\begin{split}
\frac{d\chi}{dH} \sim -S\frac{[\Delta H^2-(H-H_r)^2]}{[\Delta H^2+(H-H_r)^2]^2} -A\frac{2(H-H_r)\Delta H}{[\Delta H^2+(H-H_r)^2]^2}+Dt, 
\end{split}
\end{equation}\label{eq:eq1}
\normalsize 

where the first and second terms on the right-hand side represent the symmetric and the asymmetric contributions to the spectra, respectively. The asymmetry arises because of the electromagnetic coupling between the experimental setup and the low frequency, low amplitude magnetic modulation field. $A$ and $S$ are the asymmetric and symmetric coefficients, $H$ is the applied magnetic field and $Dt$ is linear drift term.
\footnotesize
\begin{table*}
\centering
\caption{$\alpha_{eff}^{\parallel}$ and $\alpha_{eff}^{\perp}$ represent effective damping in IP-FMR and OP-FMR, respectively, while $\alpha_{eff}^{(G+SP)}$ corresponds to the effective damping after subtracting extrinsic contributions to the damping in OP-FMR. The order of magnitude all effective damping parameters is $10^{-3}$. The error bars are within the 2\% of the given values.}
\label{tab:table2}
\begin{tabular}{c | c c c | c c c | c c c | c c c }
\hline
\hline
\textbf{} & \multicolumn{3}{ c |}{\textbf{Re (0 at\%)}} & \multicolumn{3}{ c |}{\textbf{Re (3 at\%)}} & \multicolumn{3}{ c |}{\textbf{Re (6 at\%)}} & \multicolumn{3}{ c }{\textbf{Re (12.6 at\%)}} \\
$t_{FeCo}^N$ & $\alpha_{eff}^{\parallel}$ & $\alpha_{eff}^{\perp}$ & $\alpha_{eff}^{(G+SP)}$ & $\alpha_{eff}^{\parallel}$ & $\alpha_{eff}^{\perp}$ & $\alpha_{eff}^{(G+SP)}$ & $\alpha_{eff}^{\parallel}$ & $\alpha_{eff}^{\perp}$ & $\alpha_{eff}^{(G+SP)}$ & $\alpha_{eff}^{\parallel}$ & $\alpha_{eff}^{\perp}$ & $\alpha_{eff}^{(G+SP)}$ \\
\hline
5 & 8.2 & 3.8 & 3.5 & 12.14 & 6.08 & 5.82 & 14.47 & 8.35 & 8.11 & 27.17 & 11.88 & 11.67 \\
10 & 5.91 & 2.5 & 2.07 & 9.41 & 4.63 & 4.23& 11.28 & 6.21 & 5.87 & 17.39 & 9.77 & 9.47 \\
15 & 5.7 & 2.33 & 1.78 & 7.39 & 4.08 & 3.58 & 9.72 & 5.8 & 5.35 & 15.21 & 9.3 & 8.92 \\
20 & 5.64 & 2.84 & 2.14 & 7.97 & 3.8 & 3.18 & 9.52 & 4.53 & 3.98 & 13.72 & 9.04 & 8.56 \\
30 & 6.14 & 2.93 & 1.97 & 7.83 & 4.7 & 3.85 & 9.64 & 5.54 & 4.79 & 11.69 & 8.74 & 8.10 \\
\hline
\hline
\end{tabular}
\end{table*}
\normalsize

According to the Landau-Lifshitz-Gilbert model, the linewidth is directly proportional to microwave frequency, and the proportionality constant gives the intrinsic damping parameter ($\alpha$) in the absence of inhomogeneity. However, in reality, the linewidth exhibits different contributions, \textit{e.g.} from sample inhomogeneity, spin pumping, radiative, eddy-current, and TMS. As per Arias and Mills model [\onlinecite{PhysRevB.60.7395}], TMS can be caused by lattice geometrical defects and surface-induced inversion symmetry breaking. The spin pumping contribution to the linewidth is due to the adjacent NM layer, and radiative and eddy-current contributions are due to the inductive coupling between the sample and the CPW. These other contributions will be discussed separately in the next sections. Therefore, in IP-FMR measurements, the total linewidth is a function of all contributions (\textit{i.e.}  $\mu_0 \Delta H$ = $\mu_0 \Delta H_0$ + $\mu_0 \Delta H^{(G+SP)}$ + $\mu_0 \Delta H^{rad}$ + $\mu_0 \Delta H^{eddy}$ + $\mu_0 \Delta H^{TMS}$, and is defined as,

\begin{equation}
\begin{split}
\Delta{H} = \frac{h\alpha_{eff}^{\parallel}}{g^{\perp}\mu_B\mu_0}f + \Delta{H_0}, 
\end{split}
\end{equation}\label{eq:eq2}

where $h$ is the Planck's constant, $g^\perp$ is the Landé spectroscopic splitting factor, $\mu_B$ is the Bohr magneton, $\Delta{H_0}$ is the frequency independent inhomogeneity contribution to the linewidth, and $\alpha_{eff}^{\parallel}$ = [$\alpha^{(G+SP)}$ + $\alpha^{rad}$ + $\alpha^{eddy}$ + $\alpha^{TMS}$] is known as the effective damping for the IP-FMR geometry, where $\alpha^{G}$, $\alpha^{SP}$, $\alpha^{rad}$, $\alpha^{eddy}$ and $\alpha^{TMS}$ correspond to Gilbert damping, spin-pumping damping, radiative damping, eddy-current damping and two-magnon-scattering damping, respectively. The total linewidth vs. $f$ is shown in Fig. \hyperref[fig:fig3]{3} for different Re concentrations and different FeCo thickness, and the thickness dependent effective damping is shown by red symbols in Fig. \hyperref[fig:fig6]{6}. The extracted $\mu_0 \Delta H_0$  values from IP-FMR vary in the range 2.38$-$4.29 mT for $t_{FeCo}^N$ = 5nm for all Re concentration, being smaller for $t_{FeCo}^N$ $>$ 5nm; 0.51$-$1.20 mT.

The easy axis IP-FMR Kittel equation, given below, is used to fit the experimental data [\onlinecite{PhysRev.73.155}]. The fits, shown in the insets of Fig. \hyperref[fig:fig3]{3}, give the effective magnetization ($M_{eff}$) and the IP uniaxial anisotropic field ($H_u$),

\begin{equation}
\begin{split}
f=\frac{g^\perp \mu_B\mu_o}{h}\sqrt{(H_r+H_u)(H_r+H_u+M_{eff})},
\end{split}
\end{equation}\label{eq:eq3}

In Eqs. \hyperref[eq:eq2]{2} \& \hyperref[eq:eq3]{3}, the average value of $g^\perp$ = 2.07 extracted from the OP-FMR results using asymptotic analysis of the FMR data [\onlinecite{shaw2013precise}], was used.

\subsection{OP-FMR}

\begin{figure}[b]
\includegraphics[width=9.5cm]{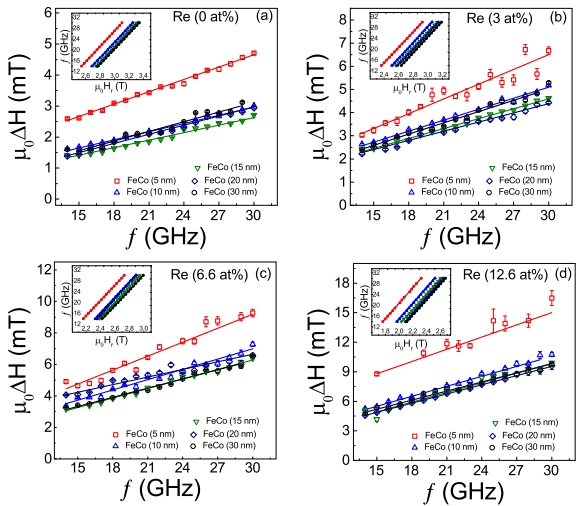}\\
\caption{$\mu_0$ $\Delta H$ vs. $f$ from OP-FMR. The insets correspond to $f$ vs. $\mu_0$ $H_r$. Open symbols and lines are experimental and fitted data, respectively.}\label{fig:fig5}
\end{figure}

Magnon scattering is the process where transitions between the uniform spin wave mode ($\textbf{k}$ = 0) and degenerate spin wave modes ($\textbf{k}$ $\neq$ 0) occur in the sample. According to Hurben and Patton [\onlinecite{hurben1998theory}], this process is strongly dependent on the wavelength of spin waves, and the coupling between $\textbf{k}$ = 0 with $\textbf{k}$ $\neq$ 0 modes are called two-magnon scattering, three-magnon scattering and so on. The level of three and higher order magnon scattering is expected to be much less than for two-magnon scattering. Therefore, it is assumed that only TMS is present in the IP-FMR results and that this contribution is minimum for the OP-FMR geometry [\onlinecite{hurben1998theory}]. Broadband OP-FMR measurements were performed where the complex transmission parameter $(S_{21})$ was recorded in field-sweep mode using VNA-technique. The used frequency range was 14$-$30 GHz. A typical result of the real and imaginary parts of the complex $S_{21}$ parameter is shown in Fig. \hyperref[fig:fig4]{4}. To analyze the data, we used the fitting model by Nembach \textit{et.al.} [\onlinecite{PhysRevB.84.054424}]. The complex $S_{21}$  function is used as follows,

\begin{equation}
\begin{split}
S_{21}(H,t) = S_{21}^0 + Dt - \frac{\chi(H)}{\tilde{\chi_o}},
\end{split}
\end{equation}\label{eq:eq4}

\begin{equation}
\begin{split}
\chi(H) = \frac{M_{eff}(H-M_{eff})}{(H-M_{eff})^2 -H_{eff}^2 -\iota \frac{\Delta{H}}{2} (H-M_{eff})},
\end{split}
\end{equation}\label{eq:eq5}

where $S_{21}^0$  is the nonmagnetic contribution to $S_{21}$ [\onlinecite{ding2004coplanar}], $Dt$ is the first order drift correction, $\tilde{\chi_o}$ is an imaginary function of frequency, film thickness and effective length of the sample on the CPW, $\mu_0 H_{eff} = hf/(g^\perp \mu_B)$, and $\mu_0 \Delta H$ is the HWHM linewidth, free from the TMS contribution. To find the precise $\mu_0 M_{eff}$ and $g^\perp$ values, considering both as fitting parameters, the OP Kittel equation can be used,

\begin{equation}
\begin{split}
f = \frac{g^{\perp} \mu_B \mu_0}{h} (H_r-M_{eff}),
\end{split}
\end{equation}\label{eq:eq6}

where $M_{eff} = M_s- H_k^\perp$, $\mu_0 H_k^\perp$ = $(2k^\perp)/(M_s t_{FeCo}^E)$ is the OP anisotropy field and $\mu_0 M_s$ is the saturation magnetization, which can be found from extrapolation of the $\mu_0 M_{eff}$ vs. $1/t_{FeCo}^E$ curve (not shown here) to $1/t_{FeCo}^E$ = 0. The fitted $\mu_0 M_s$ values are included in Table \hyperref[tab:table3]{III} for all Re concentrations, and the extracted $\mu_0 M_{eff}$  and $g^\perp$ values are given in Table \hyperref[tab:table1]{I}.

\begin{table*}
\centering
\caption{Parameters as a function of Re concentration.}
\label{tab:table3}
\begin{tabular}{c | c | c | c}
\hline
\hline
\textbf{Re (at\%)} & \textbf{$\mu_0M_S$ (T)} & \textbf{$Re(g_{\uparrow\downarrow})$ $(nm^{-2})$} & \textbf{$\alpha_{G}$ ($\times$ $10^{-3}$)}  \\
\hline
0 & 2.362(13) & 9.883(02) & 1.63(19) \\
3 & 2.183(07) & 13.791(05) & 2.82(37) \\
6.6 & 1.998(11) & 19.697(02) & 3.84(44) \\
12.6 & 1.740(03) & 18.171(03) & 7.37(26) \\
\hline
\hline
\end{tabular}
\end{table*}

\begin{table*}
\centering
\caption{Comparison of reported SMC values for different FM/NM structures}
\label{tab:table4}
\begin{tabular}{c | c | c }
\hline
\hline
\textbf{Interfaces} & \textbf{SMC ($nm^{-2}$)} & \textbf{Refs.} \\
\hline
Py/Ta & 13 & \onlinecite{ganguly2014thickness} \\
Py/Ru & 38 & \onlinecite{ganguly2014thickness} \\
Py/Pt & 26 & \onlinecite{PhysRevB.90.184401} \\
$Fe_{25}Co_{75}$/Pt & 40 & \onlinecite{PhysRevB.90.184401} \\
$Fe_{75}Co_{25}$/(Cu-Ta) & 6 & \onlinecite{schoen2016ultra} \\
Ru/$Fe_{65}Co_{35}$/Ru & 9.8 & This result \\
Ru/$(Fe_{65}Co_{35})_{93.4}Re_{6.6}$/Ru & 19.69 & This result \\
\hline
\hline
\end{tabular}
\end{table*}

\begin{figure}[b]
\includegraphics[width=9.5cm]{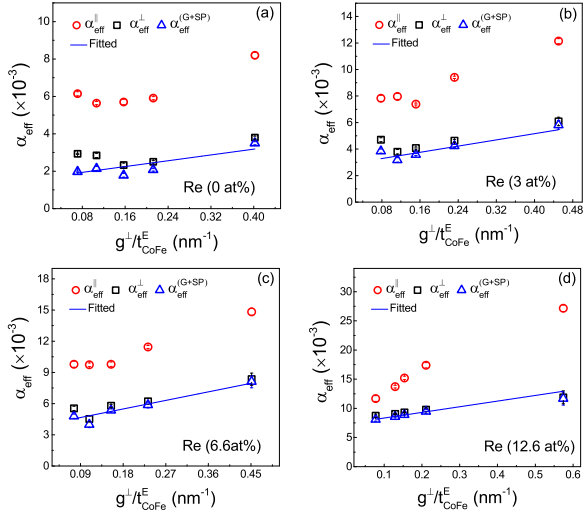}\\
\caption{$\alpha_{eff}$ vs. $g^\perp/t_{FeCo}^E$ from IP-FMR (red symbols), OP-FMR (black symbols) and $\alpha_{(G+SP)}$ (blue symbols) for all Re concentrations. The blue lines are fits according to Eq. \hyperref[eq:eq11]{11}.}\label{fig:fig6}
\end{figure}

The effective damping ($\alpha_{eff}^\perp$) can be extracted using,

\begin{equation}
\begin{split}
\Delta{H} = \frac{h\alpha_{eff}^{\perp}}{g^{\perp}\mu_B\mu_0}f + \Delta{H_0}, 
\end{split}
\end{equation}\label{eq:eq7}
where $\alpha_{eff}^\perp = \alpha^{(G+SP)} + \alpha^{rad} + \alpha^{eddy}$ is known as the OP effective damping parameter, free from the TMS contribution. The $\mu_0 \Delta H_0$ contribution to the linewidth is found to be less than 0.58 (02) mT for all samples except for 5 nm FeCo with the highest Re concentration, for which the value increased to 2.58 (07) mT. The $\mu_0 \Delta H$ vs. $f$ results are shown in Fig. \hyperref[fig:fig5]{5} for all Re concentrations and all FeCo thicknesses. A comparison of the effective damping, with ($\alpha_{eff}^\parallel$) and without TMS contribution ($\alpha_{eff}^\perp$), is presented in Fig. \hyperref[fig:fig6]{6} and in Table \hyperref[tab:table2]{II}.

In our studied films, the thickness of the Ru layer, $t_{Ru}^N$ (=3 nm)  is less than the spin diffusion length $\lambda_{Ru}^{sd}$ (=15 nm) [\onlinecite{behera2015effect}]. Therefore, significant back flow of the spin angular momentum from the Ru layer to the FeCo layer is expected. The ratio of the IP to OP effective damping ($\zeta$) can be calculated using the effective damping expression derived in Ref. [\onlinecite{PhysRevLett.114.126602}],  

\begin{equation}
\begin{split}
\zeta = \frac{\alpha_{eff}^\parallel}{\alpha_{eff}^\perp} 
\end{split}
\begin{split}
\propto \frac{\frac{1+6\eta \xi}{1+\xi}+\frac{\eta}{2(1+\xi)^2}}{\frac{1+4\eta \xi}{1+\xi}},
\end{split}
\end{equation}\label{eq:eq8}

where $\eta$ = $(\alpha_R k_F/E_F)^2$ is an interface parameter, which depends on the Rashba coefficient ($\alpha_R$), Fermi wave vector ($k_F$) and Fermi energy ($E_F$), and $\xi$ is the back flow factor. In the case of complete back flow ($\xi$ $\rightarrow$ $\infty$) $\zeta$ $\rightarrow$ 1.5, while for no back flow ($\xi$ = 0) $\zeta$ $\rightarrow$ 1.0. According to this model, $\zeta$ values in the range 1$-$1.5 are expected, while our experimentally determined $\zeta$ values for the Re-doped FeCo films are found to be greater than 1.5, which infers that $\alpha_{eff}^\parallel$ contains other contributions. The most common contribution to the effective damping enhancement in IP-FMR measurements is the TMS contribution. Therefore, in order to calculate the true value of the effective SMC, OP-FMR data devoid of TMS contributions can be used. Furthermore, extrinsic contributions, \textit{i.e.} radiative and eddy contributions, to the OP effective damping must be subtracted before determining Re-doping dependent changes in the effective SMC.

\subsection{Radiative and Eddy contributions}
According to Faraday's law, an AC voltage is generated in a metallic magnetic material when it is exposed to a time varying magnetic flux created by the microwave magnetic field and the precessing magnetization, which induces eddy-currents in the material and in the CPW. Therefore, damping due to eddy-currents in the sample is called eddy-current damping ($\alpha^{eddy}$), whereas the damping in the CPW is referred to as radiative damping ($\alpha^{rad}$). Both contributions to the effective damping depend on the wave-guide dimensions and the magnetic properties of the sample.
The eddy-current damping for the lowest order mode in FMR can be defined as [\onlinecite{schoen2016ultra},\onlinecite{PhysRevB.92.184417}],

\begin{equation}
\begin{split}
\alpha_{eddy} = \frac{C}{16} \frac{g^{\perp}\mu_B \mu_0^2 M_s (t^E)^2}{\hbar \rho},
\end{split}
\end{equation}\label{eq:eq9}

where C describes the eddy-current profile. In our case, we used C = 0.5 for all samples and $t^E$ is the total thickness of the full stack determined from XRR analysis. The $\alpha_{eddy}$ values are found to be (0.16$-$3.32) $\times$ $10^{-5}$, (0.13$-$2.81) $\times$ $10^{-5}$, (0.09$-$0.67) $\times$ $10^{-5}$ and (0.08$-$0.45) $\times$ $10^{-5}$ for the 0 at\%, 3 at\%, 6.6 at\% and 12.6 at\% Re doped samples, respectively. The first$-$last values of $\alpha_{eddy}$ within parentheses correspond to the smallest$-$largest thicknesses. The Landé $g^\perp$ factor and $\mu_0 M_s$ from the OP-FMR fitting were used in the calculations. $\rho$ is the resistivity of the full stack sample, which takes values in the range (1.243$-$0.611) $\Omega$-$\mu$m, (2.094$-$2.432) $\Omega$-$\mu$m, (1.734$-$2.290) $\Omega$-$\mu$m and (1.743$-$2.915) $\Omega$-$\mu$m for the 0 at\%, 3 at\%, 6.6 at\% and 12.6 at\% Re doped samples, respectively. The first$-$last values of $\rho$ within parentheses correspond to the smallest$-$largest thicknesses. $\alpha_{eddy}$ decreases with increasing Re doping, while $\alpha_{eddy}$ increases with an increase of the FeCo thickness.

Apart from eddy current damping, the radiative contribution to the damping also depends on the magnetic properties of the sample and the CPW parameters. The radiative damping is defined as [\onlinecite{schoen2016ultra},\onlinecite{PhysRevB.92.184417}], 

\begin{equation}
\begin{split}
\alpha_{rad} = \eta \frac{g^{\perp}\mu_B \mu_0^2 M_s t^El}{2\hbar Z_0 W},
\end{split}
\end{equation}\label{eq:eq10}

where $\eta$ is a dimensionless parameter, which describes the actual FMR mode profile in the sample. In our case, we assume $\eta$ = 0.25, considering the uniform mode of the FMR. $\mu_0 M_s$ and $g^\perp$ are used as obtained from the OP-FMR fitting. $Z_0$ = 50 $\Omega$, W = 300 $\mu$m, $t^E$ and $l$ are the CPW impedance, central width of the CPW, total thickness and length of the sample on the CPW, respectively. The $\alpha_{rad}$ values (smallest $-$ largest) are found to be (2.99$-$9.24) $\times$ $10^{-4}$, (2.61$-$8.29) $\times$ $10^{-4}$, (2.37$-$7.48) $\times$ $10^{-4}$ and (2.1$-$6.42) $\times$ $10^{-4}$ for the 0 at\%, 3 at\%, 6.6 at\% and 12.6 at\% Re doped samples, respectively. $\alpha_{rad}$ decreases with an increase of Re doping, while $\alpha_{rad}$ increases with an increase of FeCo thickness.

\begin{figure}[b]
\centering
\includegraphics[width=10cm]{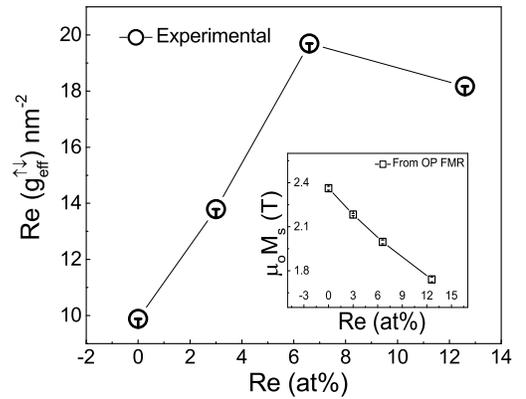}\\
\caption{The real part of the SMC for both the interface vs. Re concentration. Inset shows experimentally observed $\mu_0 M_s$ from OP-FMR.}\label{fig:fig7}
\end{figure}

\subsection{Spin pumping (SP) and Gilbert contribution}
The precessing magnetization transfers spin angular momentum to the NM layer in a FM/NM bilayer, which causes enhancement of the damping. According to the spin pumping model [\onlinecite{PhysRevB.66.224403}], the effective damping is governed by the real part of effective SMC and is expressed as,

\begin{equation}
\begin{split}
\alpha_{eff}(\alpha_{G}, \alpha_{SP})=\alpha_{G}+\frac{\mu_B}{4\pi}
\frac{Re(g^{\uparrow\downarrow}_{eff})}{ M_{s}}\frac{g^{\perp}}{t_{FeCo}^E},
\end{split}
\end{equation}\label{eq:eq11}

where $\alpha_{eff}$ is the effective damping without TMS, radiative  and eddy current contributions to the total damping, and hence only a function of the Gilbert and spin pumping contributions. The real part of the effective SMC for both the interface is obtained from thickness-dependent experimental effective damping, by linear fitting of $\alpha_{eff}$ vs. $g^{\perp}/t_{FeCo}^E$. The slope gives Re($g^{\uparrow\downarrow}_{eff}$) and the interception with the y-axis gives the Gilbert damping parameter. $\alpha_G$ is found to be 4.5 times higher with an increase of Re doping from 0 at\% to 12.6 at\%. The $\alpha_G$, Re($g^{\uparrow\downarrow}_{eff}$) and $\mu_0 M_s$ values are presented in Table \hyperref[tab:table3]{III}, and $\alpha_{eff}$ vs. $g^\perp/t_{FeCo}^E$ plots are presented in Fig. \hyperref[fig:fig6]{6} together with fits according to Eq. \hyperref[eq:eq11]{11}. 
As shown in Fig. \hyperref[fig:fig7]{7}, Re($g^{\uparrow\downarrow}_{eff}$) (per unit quantum conductance per unit area) exhibits non-monotonic behavior as a function of Re doping. The observed lowest and highest values of Re($g^{\uparrow\downarrow}_{eff}$) are found to be 9.883(02) nm$^{-2}$ and 19.697(02) nm$^{-2}$ for Re 0 at\% and Re 6.6 at\% doping, respectively. A comparison with previously reported Re($g^{\uparrow\downarrow}_{eff}$) values is presented in Table \hyperref[tab:table4]{IV}. 

According to existing theory, Re($g^{\uparrow\downarrow}_{eff}$) can be enhanced either by increasing the intrinsic SMC or by decreasing interfacial spin-memory-loss (SML) [\onlinecite{PhysRevLett.114.126602}]. A decrease of the SML with an increase of Re doping is unlikely; rather the opposite effect can be expected. The observed increase in Re($g^{\uparrow\downarrow}_{eff}$) with the increase of Re doping from 0 at\% to 6 at\% should therefore according to existing theory be linked to the increase of the intrinsic SMC. However, this would require an increase of the number of quantum mechanical conductance channels in the NM layer, which again seems unlikely considering that the Re doping takes place in the FM layer. As we will argue below, the explanation for the increase of Re($g^{\uparrow\downarrow}_{eff}$) can instead be found in the influence of the spin-orbit coupling (SOC) on the effective SMC. 

The intrinsic SMC calculated using the existing theory [\onlinecite{PhysRevB.66.224403},\onlinecite{brataas2004spin}] is only well defined without SOC. It is nevertheless possible from theory to identify an effective SMC [Re($g^{\uparrow\downarrow}_{eff}$)] by calculating the interface-enhanced damping parameter. Theoretical calculations for Pt/FeCo varying the concentration of Co show that the magnitude of Re($g^{\uparrow\downarrow}_{eff}$) exhibits a non-monotonic behaviour with Co-concentration and that Re($g^{\uparrow\downarrow}_{eff}$) can be increased by a factor of two in the presence of interfacial SOC on the NM side [\onlinecite{PhysRevB.98.174412}]. This change in Re($g^{\uparrow\downarrow}_{eff}$) results from the partially filled $d$-bands and half-filled $s$-band of Pt (5d$^9$6s$^1$). The interaction between the $d$-bands of FeCo and Pt near the Fermi level results in the non-monotonic Co-concentration dependent behaviour of Re($g^{\uparrow\downarrow}_{eff}$). Similar calculations for Au(5d$^{10}$6s$^1$)/FeCo show that  Re($g^{\uparrow\downarrow}_{eff}$) is less influenced by SOC due to the mainly $s$-electron character of the conduction channel in Au. In addition, calculations of the Re($g^{\uparrow\downarrow}_{eff}$) for Pt/Py indicated an increase by 25\% in the presence of interfacial SOC, while no significant change was observed for Cu(3d$^{10}$4s$^1$)/Py [\onlinecite{PhysRevLett.113.207202}]. For comparison, the experimentally observed enhancement in Re($g^{\uparrow\downarrow}_{eff}$) for Ru(4d$^7$5s$^1$)/FeCo [cf. Fig. \hyperref[fig:fig7]{7}] can also be attributed to interfacial SOC. As the Re-concentration increases, the density of states at the Fermi level increases, which will also modify the interaction between the $d$-bands of FeCo and Ru and the interfacial SOC in the Ru layer. The interpretation for low Re-concentrations is that the interfacial SOC in the Ru layer increases with increasing Re-concentration. This interpretation might also explain the non-monotonic behaviour of Re($g^{\uparrow\downarrow}_{eff}$) with increasing Re-concentration. However, the decrease of Re($g^{\uparrow\downarrow}_{eff}$) at higher Re doping may be due to SML, since an increase of the SML for large enough Re-concentration cannot be ruled out [\onlinecite{PhysRevLett.114.126602}].

\section{\label{sec:four}Conclusion}
The dynamics properties of Re-doped (0$-$12.6 at\%) polycrystalline $Fe_{65}Co_{35}$ thin films with Ru as capping and seed layers have been investigated using room temperature IP- and OP-FMR measurements. Comparison of the effective damping in IP- and OP-FMR indicates that the IP damping is affected by TMS, which overestimates the real part of the effective spin mixing conductance. The thickness dependent results for the effective damping, after subtracting the radiative and the eddy-current damping contributions, indicate that the enhancement of the damping is due to the spin pumping contribution. By further analyses of the OP-FMR results, free from TMS, radiative and eddy-current contributions, a non-monotonic dependence of the real part of the effective spin-mixing conductance on Re concentration is found. The increase of Re($g^{\uparrow\downarrow}_{eff}$) with increasing doping from 0 at\% to 6 at\% Re is tentatively explained by a corresponding increase of the interfacial SOC in the Ru layer. Apart from this, an enhancement of the bulk Gilbert damping is found with an increase in Re doping, while $\mu_0 M_s$ decreases with increasing Re concentration. This study opens an entirely new direction of tuning the spin-mixing conductance in magnetic heterostructures by doping the ferromagnetic layer, thus providing a method for optimizing the design of spintronic devices.

\section*{Acknowledgement}
This work is supported by the Swedish Research Council (VR; grant no 2017-03799), Olle Engkvist Byggmästare (project number 182-0365), and by the Marie Curie Action ‘Industry-Academia Partnership and Pathways’ (ref. 612170, FP7-PEOPLE-2013-IAPP).
\bibliography{ref}
\end{document}